\documentclass[12pt,a4paper]{article}

\usepackage{amsmath,latexsym,amssymb}
\usepackage{hyperref}
\usepackage{color}

\newcommand{\be}{\begin{equation}}
\newcommand{\ee}{\end{equation}}
\newcommand{\bea}{\begin{eqnarray}}
\newcommand{\ena}{\end{eqnarray}}

\newcommand{\nb}{\nonumber}

\renewcommand\o{\omega}

\renewcommand\t{\ensuremath{\theta}}

\newcommand\x{\ensuremath{\times}}

\newcommand\m{\ensuremath{\mu}}

\newcommand\n{\ensuremath{\nu}}
\newcommand\C{{\ensuremath{\cal C}}}
\newcommand\M{{\ensuremath{\cal M}}}
\newcommand\K{{\ensuremath{\cal K}}}

\newcommand\diag{\text{diag}}
\newcommand\PM[1]{\begin{pmatrix}#1\end{pmatrix}}
\newcommand{\de}{\partial}

\newcommand{\ben}{\begin{displaymath}}
\newcommand{\een}{\end{displaymath}}
\newcommand{\ba}{\begin{eqnarray}}
\newcommand{\ea}{\end{eqnarray}}
\newcommand{\ban}{\begin{eqnarray*}}
\newcommand{\ean}{\end{eqnarray*}}
\newcommand{\bs}{\begin{split}}
\newcommand{\es}{\end{split}}

\title{Gravity Modification with Yukawa-type Potential:\\ 
Dark Matter and Mirror Gravity}

\author{Zurab Berezhiani,\!\footnote{
\texttt{berezhiani@fe.infn.it}} ~ 
Fabrizio Nesti,\!\footnote{
\texttt{fabrizio.nesti@aquila.infn.it}} ~ 
Luigi Pilo,\!\footnote{
\texttt{luigi.pilo@aquila.infn.it}} ~ 
Nicola Rossi\footnote{
\texttt{nicola.rossi@aquila.infn.it}}  
\\ \\
{\small\it Dipartimento di Fisica, Universit\`a di L'Aquila,  
      67010 Coppito AQ, and  } 
\\[-.5ex]
{\small\it INFN, Laboratori Nazionali del Gran Sasso, 67010 Assergi AQ, Italy} }

\begin{document}
\date{}
\maketitle


\begin{abstract}
  \noindent The nature of the gravitational interaction between
  ordinary and dark matter is still open. Any deviation from
  universality or the Newtonian law also modifies the standard
  assumption of collisionless dark matter. On the other hand,
  obtaining a Yukawa-like large-distance modification of the
  gravitational potential is a nontrivial problem, that has so far
  eluded a consistent realization even at linearized level. We propose
  here a theory providing a Yukawa-like potential, by coupling
  non-derivatively the two metric fields related respectively to the
  visible and dark matter sectors, in the context of massive gravity
  theories where the local Lorentz invariance is broken by the
  different coexisting backgrounds. This gives rise to the appropriate
  mass pattern in the gravitational sector, producing a healthy theory
  with the Yukawa potential. Our results are of a special relevance in
  the scenario of dark matter originated from the mirror world, an
  exact duplicate of the ordinary particle sector.

\end{abstract}
   
\vfill

\vspace*{1cm}

\newpage

\section{Introduction}
The problem of obtaining a Yukawa-like potential in a consistent
theory of gravity is a nontrivial task and attempts in this direction
date back to 1939 when Fierz and Pauli (FP) added a mass term $m$ to
the Lorentz-invariant action of the free spin-2 graviton~\cite{FP}.
Unfortunately, the resulting FP theory of massive gravity is unfit to
be a consistent modification of GR because of the van
Dam-Veltmann-Zakharov (vDVZ) discontinuity \cite{DIS}: also in the
limit $m \to 0$ the bending of light is $25\%$ off the extremely
precise experimental limit.

A further, theoretical, problem is that the fine-tuning needed to
single out the ghost-free action at linearized level is spoiled by
interactions and a sixth ghost-like mode starts to propagate, making
the whole theory unstable~\cite{boul_des} and unpredictable
\emph{below} some (unacceptably large) distance scale. The problem was
reexamined in the framework of effective field theories realising that
the reason behind the misbehavior of FP massive gravity is strong
coupling of the scalar sector~\cite{NGS}.

It has been shown that the sickness of the FP theory has its roots in
the Lorenz invariance \cite{ghostcond}.  Indeed, retaining only
rotational invariance, one can avoid the vDVZ discontinuity and the
propagation of ghost-like states~\cite{rubakov,dubos} (for a different
approach, see~\cite{kb}). In these models, gauge invariance is broken
by Lorentz-breaking mass terms, and the gauge modes that would start
propagating, acquire a well behaved kinetic term, or do not propagate
at all. What happens is that via Lorenz-breaking one can cure the
discontinuity problem in the ``spatial'' sector while avoiding
ghost-like propagating states.

In the context of bigravity \cite{Isham}, a suitable realization of a
Lorenz-breaking (LB) massive phase of gravity can be
obtained~\cite{pilo,Blas:2007zz}. In addition to our metric field $g_{1\mu\nu}$
and normal matter minimally coupled to $g_{1\mu\nu}$, described by the
Lagrangian ${\cal L}_1$ (sector 1 in the following), one introduces a
second metric tensor $g_{2\mu\nu}$ related to a hidden sector 2 (dark
matter) with Lagrangian ${\cal L}_2$. Therefore, the visible and dark
components are associated to separate gravitational sectors. The
action of this theory consists of an Einstein-Hilbert term for each
metric, plus an interaction term $V$
\be
S = \int d^4 x \, \left[\sqrt{g_1} \, \left(M_1^2 R_1  + {\cal L}_1 \right)+ 
\sqrt{g_2} \, \left(M_2^2 \, R_2  + {\cal L}_2 \right) + 
\epsilon^4 \, \left(g_1 g_2 \right)^{1/4} V(X) \right] ,
\label{bg}
\ee 
where $M_{1,2}$ are the ``Planck'' masses of the two sectors and
$\epsilon$ is some small mass scale which essentially will define the
graviton mass through a see-saw type relation $m_g \sim
\epsilon^2/M_P$, $M_P$ being the experimental Planck mass (of course
related to $M_{1,2}$). The metric determinants are denoted $g_1$ and
$g_2$, and the interaction potential $V$ among the two metrics is
assumed to be non-derivative, so that it can always be taken as a
scalar function of $X^\mu_\nu = g_1^{\mu \alpha} {g_2}_{\alpha
  \nu}$. The invariance under diffeomorphisms is not broken; local
Lorentz invariance, on the other hand, is spontaneously broken because
in general there is no local Lorentz frame in which the two metric
tensors $g_{1\mu\nu}$ and $g_{2\mu\nu}$ are proportional. Nonetheless,
because each matter sector is minimally coupled to its own metric, the
weak equivalence principle is respected and the breaking of local
Lorentz invariance is transmitted only gravitationally. Once a flat
rotationally invariant background is found, in the weak field limit a
Lorentz breaking mass term for the gravitational perturbations emerges
in a natural way from the expansion of the interaction term the
action~\cite{pilo}.

One should point out that when $V$ is absent the gauge symmetry is
enlarged, because one can transform $g_{1\mu\nu}$ and $g_{2\mu\nu}$ by
using two independent diffeomorphisms. On the other hand when $V$ is
turned on the symmetry is reduced to the the common (diagonal)
diffeomorphism group, corresponding to general covariance. As a result
a massless graviton is always present, besides the massive
excitations. As shown in~\cite{pilo}, in the Lorentz breaking phase
only tensors propagate, in particular in the vector and scalar sectors
no time derivatives are present. The Newtonian potential is modified
in the infrared, but the modification is not Yukawa-like. In fact, at
linearized level, the deviation from a $1/r$ potential is a linearly
growing term~\cite{dubos,pilo}.\footnote{Such a term clearly breaks
  perturbativity at some large distance $r>r_{\text{IR}}$, but
  remarkably this behavior is cured by the non-perturbative
  treatment~\cite{pilo1} showing that the linear term is replaced by a
  non-analytic term $r^\gamma$.}

In order to have a massive phase with a Yukawa-like potential, 
bigravity must then be enlarged.  In this paper we generalize the
above construction and show that one can use a further rank-2 field
$g_{3\mu\nu}$ as a Higgs field to achieve the Yukawa potential.  The
size of the fluctuations of $g_{3\mu\nu}$ is controlled 
by a 3rd "Planck" mass $M_3$ entering in its EH action. 
We will show that, in the limit $M_3 \gg
M_{1,2}$, $g_{3\mu\nu}$ can be consistently decoupled and one is left
with an effective bigravity theory with a Yukawa-like component of the
gravitational potential.  The tensor $g_{3\mu\nu}$ plays the role of a
symmetry-breaking field, communicating the breaking of Lorentz
invariance to $g_{1\mu\nu}$ and $g_{2\mu\nu}$ and thus introducing 
Lorentz-breaking mass terms to their fluctuations. 
The resulting phase of gravity features a Yukawa-modified static
potential while still avoiding any propagation of ghosts and the vDVZ
discontinuity, and this result survives in the limit of decoupling $g_3$.

This modification of gravity at large distances can then open new
possibilities for the nature of dark matter. In the present paradigm
the visible matter amounts only to about $4\%$ of the present energy
density of the Universe while the fraction of dark matter is about 5
times bigger. Cosmological observations are consistent with the
hypothesis of cold dark matter. On the other hand, the situation at
the galactic scale is rather different. According to the CDM paradigm
cold and collisionless dark matter is distributed, differently from
the visible matter, along the galactic halos and is responsible for
the anomalous behaviour of galactic rotational curves. However, in CDM
the curves obtained using N-body simulations do not reproduce the
observed rotational curves of small galaxies~\cite{sal1,sal2}. The
implicit assumption behind this scenario is that gravitational
interaction between matter and dark matter is universal and Newtonian.
Relaxing these hypotheses may radically modify our view and
phenomenological modelling.

One of the intriguing possibilities is to consider dark matter as
matter of a hidden gauge sector which is an exact copy of the ordinary
particle sector, so that along with the ordinary matter: electrons,
nucleons, etc.\ the Universe contains also the mirror matter as mirror
electrons, mirror nucleons, etc. with exactly the same mass spectrum
and interaction properties. Such a parallel sector, dubbed as mirror
world~\cite{mirror}, can have many interesting phenomenological and
cosmological implications (for a review, see~\cite{alice}). In
particular, the baryon asymmetry in both sectors can be generated via
out-of-equilibrium $B\!-\!L$ and $CP$ violating processes between
the ordinary and mirror particles \cite{BB-PRL} and it can naturally
explain the proportion between the visible and dark matter fractions
in the Universe. Such processes can be induced by some very weak
interaction between ordinary and mirror sectors. The very same
interaction can also induce mixing terms between neutral particles of
the two sectors, as e.g. kinetic mixing for photons \cite{Holdom} or
mass mixing in the case of the neutrinos and neutrons \cite{FV}.

Mirror matter, dark in terms of ordinary photons, is couples with
ordinary matter through gravity and can be a viable candidate for dark
matter. As it was shown in \cite{BCV}, the cosmological observations
on large scale structure and CMB are consistent with the mirror dark
matter picture. However, the essential problem emerges at the galaxy
scales. It is difficult to understand how the mirror matter, being
collisional and dissipative as normal matter, could produce extended
galactic halos and thus explain the galactic rotational curves.

In this paper we show that this difficulty is overcome if the
mirroring is extended also to the gravitational sector as encoded in
the action (\ref{bg}), normal and mirror matter having separate
gravities generated respectively to the metric fields $g_{1\mu\nu}$
and $g_{2\mu\nu}$. A healthy Yukawa modification of the gravitational
potential appears when Lorentz breaking is induced by the third metric
$g_{3\mu\nu}$ (whose "Planck" mass $M_3$ is eventually taken to be
much larger than the ordinary Planck mass $M_P$). Explicitly, the
potential felt by a test particle of the type 1 (normal matter) at the
distance $r$ from a source is
\begin{equation} \label{bigr}
  \phi(r)= \frac{G}{2r} \left[ \left( m_1 + m_2  \right) +
  \left( m_1 - m_2 \right) \, e^{-r/r_m } \right] ,
\end{equation}
where $G$ is the Newton constant, $m_1$ and $m_2$ are respectively the
masses of the visible (type 1) and mirror (type 2) matter sources, and
$r_m=m^{-1}$ is a Yukawa length scale. Hence, at small distances $r
\ll r_m$, gravitational force between two sectors is not universal:
normal and mirror matter effectively do not interact gravitationally .
At distances $r \ll r_m$ a normal test mass interacts only with $m_1$
through the ordinary Newton potential. However, at large distances $r
\gg r_m$ gravity becomes universal and Newtonian, a test particle
feels both ordinary and dark matter and attributes to the source a
total mass $(m_1 + m_2)$ with an effectively \emph{halved} Newton
constant $G/2$. The main result of this work is to reproduce the
potential (\ref{bigr}) in a consistent model of gravity.

\pagebreak[3]

This scenario can have interesting astrophysical implications. One can
show~\cite{curves} the galactic rotational curves are reproduced even
if dark matter\footnote{For bigravity inspired interpretation of dark matter see \cite{Banados:2008fj}.} has a similarly "clumped" distribution as normal
matter, as it is expected from its dissipative character.

The paper is organized as follows: in section~2 we review the
linearized analysis of bigravity theories that will be used as
building block for our model, in particular both the Lorentz breaking
and Lorentz invariant phases are discussed. In section~3 we introduce
the model and show how a Yukawa potential arises in the limit when the
additional metric is decoupled. In section~4 we discuss our findings.
Finally, in appendix A the general expression of the Yukawa-like
potential is given, and appendix B contains the detailed expressions
for the graviton mass matrices and an example of the interaction
potential having all the required features.

\section{Bigravity: A review of the Linearized Analysis }

In bigravity generically one can
find bi-flat $SO(3)$ preserving vacuum solutions~\cite{pilo}:
\be
\begin{split}
& \bar   g_{1\mu\nu} = \eta_{\mu\nu} = {\rm diag}(-1,1,1,1) \\  
& \bar   g_{2\mu\nu} =\hat{\eta}_{\mu\nu} = \omega^2 {\rm diag}(-c^2,1,1,1) ;
\end{split}
\label{vac}
\ee 
we have set the speed of light in our sector (1) to one in natural units, whereas 
$c$ is the speed of light in the hidden sector 2 and $\omega$ is a relative
constant conformal factor.  Once $V$ is given, $c$ and $\omega$ can be
computed by solving the equations of motion following from~(\ref{bg}),
and if $c \neq 1$ Lorentz symmetry is broken.  Consider the
linearized theory obtained by expanding the total action (\ref{bg}) at
quadratic level in the metric perturbations around the bi-flat
background ({\ref{vac}):
\be
  g_{1\mu\nu} =  \eta_{\mu\nu} + h_{1\mu\nu} \, ,\qquad   
g_{2\mu\nu} =  \hat{\eta}_{\mu\nu} + \omega^2h_{2\mu\nu} \ . 
\ee
The gravitational perturbations $ h_{1\,\mu\nu}$ and $ h_{2\,\mu\nu} $
interact with matter 1 and 2 through their conserved EMTs,
respectively $ T^{\mu\nu}_1 $ and $ T^{\mu\nu}_2 $.  Since the
background preserves rotations, it is convenient to decompose the
perturbations $h_{a\m\n}$ ($a=1,2$) according to irreducible $SO(3)$
representations
\be
\begin{split}
& {h_a}_{00} = \psi_a \,, \qquad {h_a}_{0 i} = {u_a}_i + \de_i v_a\,, \\[.5ex]
&  {h_a}_{ij} = {\chi_a}_{ij} + \de_i {S_a}_j + \de_j {S_a}_i + 
   \de_i \de_j \sigma_a + 
\delta_{ij} \, \tau_a \,.\\[-.5ex]
\end{split}
\label{eq:decomp}
\ee
For each perturbation ${h_a}_{\mu \nu}$ one has a gauge invariant transverse traceless  
tensor $\chi_{a ij}$, two vectors and  four scalars. The quadratic 
Lagrangian ${\cal L}$ reads
\bea
\label{lag}
\mathcal{L} &=& \mathcal{L}_{\text{kin}} + \mathcal{L}_{\text{mass}}+ \mathcal{L}_{\text{src}} \,,
\\
\mathcal{L}_{\text{kin}}\!&=&\!
\frac14 \chi_{ij}^t \K(\C^2 \Delta - \de_t^2)\chi_{ij}
- \frac12 w_i^t \K \Delta  w_i+
\\
\!&&\!{}+ \phi^t \K  \Delta \,  \tau 
- \frac12 \tau^t \K \left(\C^2 \Delta -3 \de_t^2  
  \right) \tau .
\label{kin}
\ena 
We have introduced a compact notation for the fluctuations: $h_{\m\n}
\!=\! (h_{1\m\n},h_{2\m\n})^t$, $\chi_{ij} \!=\!
(\chi_{1ij},\chi_{2ij})^t$ and the following 2\x2 matrices: $\C =
\diag (1, c) $, $\K = M_1^2 \, \diag (1, \kappa)$ and $\kappa = M_2^2/M_1^2 \o c$. 
In the kinetic term coming from the expansion of EH terms, the fluctuations
enter only through the gauge invariant combinations ${\chi_a}_{\mu \nu}$, $w_i = u_i - \de_t
S_i $ and  $\phi = \psi - 2 \de_t v + \de_t^2 \sigma$. Finally, $\mathcal{L}_{\text{src}}$
describes the gravitational coupling to matter  conserved sources
\be
\label{src}
\mathcal{L}_{\text{src}} = 
T^t_{0i}\, \C^{-1} W_i - T^t_{00} \frac{\C^{-3}}{2} \phi 
-  T^t_{ii} \frac{\C}{2} \tau  - T^t_{ij} \frac{\C}{2} \chi_{ij} \,.
\ee 
Clearly $\mathcal{L}_{\text{kin}}$ and $\mathcal{L}_{\text{src}}$ are
gauge invariant. The mass term
$\mathcal{L}_{\text{mass}}$ is produced by the expansion of the
interaction potential $V$. For the bi-flat background~(\ref{vac}) the mass term
$\mathcal{L}_{\rm m}$ has the following form
\be
\label{mass}
\mathcal{L}_{\text{mass}}
= \frac{\epsilon^4}{4} \Big( h_{00}^t \M_0 h_{00}
     +2 h_{0i}^t \M_1 h_{0i}  -  h_{ij}^t 
   \M_2 h_{ij} +
h_{ii}^t \M_3 h_{ii} 
     - 2  h_{ii}^t \M_4 h_{00} \Big)\,
\ee
and the explicit value of the mass matrices can be easily computed for
any given $V$. 

It is however crucial to realize that due to linearized gauge
invariance the mass matrices have the following property~\cite{pilo}
\be
\label{condM}
\M_{0,1,4}\begin{pmatrix}1 \\ c^2  \end{pmatrix}=0\,, 
 \qquad\M_{1,2,3} \begin{pmatrix} 1  \\ 1  \end{pmatrix}=0\,, \quad 
 \quad\M_{4}^t \begin{pmatrix} 1  \\ 1  \end{pmatrix}=0 \;.
\ee
Thus general covariance forces the mass matrices  to be at most of rank one.

\paragraph{Lorenz-Invariant (LI) phase.}

In this case an FP graviton mediates Yukawa-like potential. 
Indeed, when $c=1$, two conditions in~(\ref{condM}) coincides, allowing a non-zero $\M_1$ and 
all mass matrices are rank one and proportional:
\be
\M_0 = \lambda_0 {\cal P}, \quad 
\M_1 = \M_2 = \lambda_2 {\cal P}, \quad   \quad 
\M_3 = \M_4 = (\lambda_2 + \lambda_0) {\cal P} \; ,
\quad  {\cal P} = \begin{pmatrix} 1 & - 1 \\
  -1 & 1  \end{pmatrix} .
\ee
After introducing a canonically normalized graviton field $h^{(c)}= {\mathcal K}^{1/2} h$,
the mass matrices can be diagonalized by a rotation of angle $\vartheta$ with $\tan
\vartheta = \kappa^{1/2}=M_2/\o M_1$; the spectrum consists of a 
massless and a massive graviton.  This latter has a standard
Lorentz-Invariant mass term and to avoid ghosts one has to choose
$\lambda_0 = 0$, as a result the mass term has the Pauli-Fierz form.
Both matter sectors interact with the massless graviton and mediates
a force with a standard Newtonian potential. The massive graviton on the other hand mediates a  Yukawa-like force and thus modifies the static potential at scales larger than $m^{-1}$,
where $m = \epsilon^2 \lambda_2^{1/2} |\sin\vartheta| /M_1$ is the graviton mass.

\pagebreak[3]

In the most interesting case, when $M_1=M_2=M$, $\omega=1$ and so
$\tan \vartheta = 1$, a static potential for a test particle of type
1, generated by point-like sources of mass $m_1$ (type-1) and $m_2$
(type-2) at the same point, is given by
\begin{equation} \label{eq_matpotsym}
\phi_{1\text{matter}}(r)=  \frac{1}{32 \pi M^2 } 
\left[ \left( m_1 + m_2  \right) +
  \frac{4}{3} \, e^{-m \, r}
  \left( m_1  - m_2 \right) \right].
\end{equation}
Therefore, the presence of the massive gravity state mediating 
a Yukawa-like force makes the effective Newton constant distance 
dependent: the Newton constant measured experimentally via 
the gravitational interaction between type-1 test bodies  
at small distances $r \ll m^{-1}$ should be identified with  
$G=G_{UV} = 7/96\pi M^2$. At large distances $r \gg m^{-1}$
it effectively becomes $G_{IR}= 1/32\pi M^2=3G/7$ 
and is universal for both type-1 and type-2 matter.  
On the other hand, at  $r \ll m^{-1}$ 
the force between the type-1 and type-2 test masses 
is repulsive, with an effective Newton constant $G^{12}_{UV} = - G/7$; this is an indication 
of  instability of the theory. However, more serious problem 
is the vDVZ discontinuity. The static potential felt by a
photon is 
\begin{equation}
  \phi_{1\text{light}}(r)=   \frac{1}{32 \pi M^2 }
  \left[ \left( m_1 +m_2 \right) +  e^{- m r} \left( m_1  - m_2 \right) \right]. 
\end{equation}
Therefore, for the light bending at distances $r \ll m^{-1}$ we have 
$G_{\rm light} = 1/16\pi M^2$, and thus $G/G_{\rm light} = 7/6$. 
This discrepancy  is somewhat milder than in the FP theory 
where we have $G/G_{\rm light} = 4/3$; anyway 
the deviation from the GR prediction  $G/G_{\rm light} = 1$ 
is unacceptably large and it 
is clearly excluded by the post-Newtonian gravity tests \cite{will}. 

The problems can be softened if the two sectors are not symmetric, 
$M_1\neq M_2$ and  is a small enough mixing angle. 
In this case, the static potential respectively  a massive test  
particle and  for a  photon of the type 1 reads 
\begin{equation} \label{eq_matpot}
\phi_{1\xi}(r) =  \frac{\cos^2\vartheta}{16 \pi M_1^2 }
\left[ \left( m_1 + m_2  \right) +  \xi \, e^{-m \, r}
  \left( m_1 \, \tan^2 \vartheta - m_2 \right) \right], 
\end{equation}
where $\xi=4/3$ for a massive particle and $\xi=1$ for light.
Therefore, at small distance ($ r \ll m^{-1}$) the Newton ``constant'' is $G_{UV}= G
(1+ 4/3 \tan^2 \vartheta)$, at large distance ($r \gg m^{-1}$) it tends to
$G$.  

At small distances,  the ratio of (\ref{eq_matpot}) 
defines the post-Newtonian parameter
$\delta $:
\begin{equation}
  \delta = \lim_{m \to 0} \left[ \frac{\phi_{1\xi}(\xi=4/3)}{\phi_{1\xi}(\xi=1)} 
  \right]_{m_2 = 0}= 1+ 
  \frac{1}{3} \,\sin^2 \vartheta   \; .
\end{equation}
For GR $\delta = 1 $ and  the current light bending experiments constrain $ \delta$ to be in the range $ \delta =1.0000 \pm 0.0001 $~\cite{will}. In the 
limit of vanishing graviton mass the well known vDVZ
discontinuity~\cite{DIS} of Pauli-Fierz massive gravity emerges. In our case
the mixing angle $ \vartheta $ controls the size of the discontinuity.

When $M_2 \gg M_1$, we have $\vartheta\to \pi/2$ and $\delta=4/3$,
unacceptably large.  In this limit sector 2 is very weakly coupled,
and the discontinuity is mainly shifted to sector 1, that approaches a
normal Fierz-Pauli massive gravity.

Conversely when $M_2 \ll M_1$ we have $\vartheta\to 0$ and the
discontinuity is shifted to sector 2; $h_+$ and $h_-$ almost coincide
with $ h_1 $ and $ h_2 $ and gravity is stronger in sector 2. The
experimental bound on $\delta$ translates into $ \vartheta \simeq
0.02$, that amounts to roughly $ M_2 \simeq \vartheta M_1$.  
In this case, if $ m_2 $ is interpreted as dark matter, it gives a
sizable contribution, increasing the gravitational force in the
region $ r \gtrsim m^{-1} $.  Notice incidentally that for small $ r $,
the potential is repulsive. This result
contradicts observations in the gravitationally bounded systems as cluster
and galaxies, for this reason is ruled out.

\paragraph{Lorenz-Breaking (LB) phase.}
In this phase, $c \neq 1$, conditions~(\ref{condM}) imply that
$\M_1$=0 and for other masses one has
\vspace*{-.5ex}
\be
\M_0 = \lambda_0  \, {\cal C}^{-2} \, {\cal P} \,  {\cal C}^{-2} \, , 
\quad 
\M_{2,3} =\lambda_{2,3} \,  {\cal P} 
 \, , \quad  \M_4 = \lambda_4 \, {\cal P}  \, {\cal C}^{-2} \; .
\label{eq:masses}
\ee
In this situation all the scalar and vector perturbations become
non-dynamical~\cite{pilo}. The vanishing of $\M_1$ in the LB phase is
the reason behind the absence of  ghosts or tachyons appear, gravitons
are the only propagating states. However, this is also the reason
behind the absence of Yukawa-like gravitational potential.
The resulting modification was studied in detail in \cite{pilo1} both at linear 
and non-linear  level.

\section{Effective Higgs Phase}

In order to find a phenomenologically viable Yukawa phase, we
introduce one more rank-2 field $ g_3 $ which couples both metrics $ g_1
$ and $ g_2$ and triggers LB:\footnote{In principle, any tensor condensate e.g.
  emerging via a strongly coupled hidden gauge sector can be also used
  for inducing the Lorentz-breaking background \cite{Kancheli}. }
\bea \label{tri}
 S\!\! &=&\!\! \int\! d^4 x \, \left[\sqrt{g_1} \, \left(M^2 R_1  + 
{\cal L}_1 \right)+ 
\sqrt{g_2} \, \left(M^2 \, R_2  + {\cal L}_2 \right) +  
M_3^2 \, \sqrt{g_3} R_3  
\right. \nb\\
&& \qquad\quad\left. {}+ \epsilon^4 \, \left(g_1 g_2 g_3 \right)^{1/6}  V(g_1,g_2, g_3) \right]  \; .
\ena
The only non-trivial tensors that can be formed are: $ X_{12} =
g_1^{-1} g_2 $, $ X_{13} = g_1^{-1} g_3 $, $ X_{23} = g_2^{-1} g_3 $,
that satisfy the identity $X_{13}=X_{12}X_{23}$. Therefore $V$ can be
taken as a scalar function of two of them.  

We have also introduced  in~(\ref{tri}) a discrete symmetry under the
exchange 1$\leftrightarrow$2, so that the potential $V$ is symmetric
and the two sectors 1, 2 have equal Planck masses $M$. The third
Planck mass will be eventually taken to infinity,
$M_3\gg M$, and the fluctuations of the third field will be
effectively decoupled.

The first step is to find a suitable background. As for bigravity, we
look for LB flat solutions of the form 
\be
\begin{split}
&\bar g_{1 \mu \nu} = \bar g_{2 \mu \nu} = \eta_{\mu \nu} ={\rm diag}(-1,1,1,1) 
\\[.7ex]
&\bar g_{3 \mu \nu} = \hat \eta_{\mu \nu} = \omega^2 \, {\rm diag} \left(-c^2, 1,1,1 \right) ,
\end{split}
\label{vac3}
\ee
so that $ \bar{X}_{12} = \mathbb{I} $ and $ \bar{X}_{13} = \eta^{-1}
\hat{\eta} $. The background (\ref{vac3}) is a solution of the
equations of motion (EOMs) if
\begin{eqnarray}
  V \mathbb{I} + 6 \frac{\partial V}{\partial X_{21}} X_{21} + 
  6 \frac{\partial V}{\partial X_{31}} X_{31} &=& 0 \nb\\
  V \mathbb{I} + 6 \frac{\partial V}{\partial X_{12}} X_{12} + 
  6 \frac{\partial V}{\partial X_{32}} X_{32} &=& 0 \nb\\
  V \mathbb{I} + 6 \frac{\partial V}{\partial X_{13}} X_{13} + 
  6 \frac{\partial V}{\partial X_{23}} X_{23} &=& 0\, ,
\end{eqnarray}
where $X_{ba}^{\vphantom{-1}}=X_{ab}^{-1}$. Using the 1$\leftrightarrow$2 exchange symmetry of the EOMs, the symmetric form of the ansatz  (\ref{vac3}) and the   the identity
\begin{equation}
  \frac{\partial V}{\partial X_{ab}} X_{ab}=
  -\frac{\partial V}{\partial X_{ba}} X_{ba} \,,
\end{equation}
we have $\partial V/\partial X_{12}=0$ and the EOMs reduce to
\begin{equation}
\label{eq:const}
  V = 0\,,\qquad 
  \frac{\partial V}{\partial X_{13}}=0\,.
\end{equation}
These are three independent equations for the two parameters $\omega$
and $c$, thus one fine-tuning is needed for  (\ref{vac3}) to be a 
solution.  This fine tuning is analogous to the cosmological constant
in standard GR, and can be easily realized for instance by introducing
a cosmological constant in sector 3.

\medskip

Once a background solution is found, one can study small fluctuations around it, 
defined by
%
\be
 g_{1\mu\nu} =  \eta_{\mu\nu} + h_{1\mu\nu} \, , \qquad 
g_{2\mu\nu} =  \eta_{\mu\nu} + h_{2\mu\nu} \, , \qquad
 g_{3\mu\nu} =  \hat \eta_{\mu\nu} + \omega^2 h_{3\mu\nu} \,. 
\ee
The structure of the quadratic Lagrangian for the fluctuations is the
same as in (\ref{lag})-(\ref{src}) except that now the tensor, vector,
scalar and source fields all have 3 components, $ h_{\mu\nu} = (h_{1
  \,\mu\nu},h_{2 \,\mu\nu},h_{3 \, \mu\nu} ) $.  Also,
\begin{equation}
   \K = {\rm diag}(M^2,M^2,M_{3}^2/\omega^2c)\,, \qquad
   \C = {\rm diag}(1,\,1,\,c) \; , 
\end{equation}
and $ \M_i$ are 3$\times$3 matrices entering the usual
mass Lagrangian:
\begin{equation} 
  \mathcal{L}_{\text{mass}} =
 h_{00}^t \M_0 h_{00} + 
  2 h_{0i}^t \M_1 h_{0i} - h_{ij}^t \M_2 h_{ij} + 
  h_{ii}^t \M_3 h_{ii} -2 h_{ii}^t \M_4 h_{00}\,.
\end{equation}

Diagonal diffeomorphisms invariance constrains the form of these
matrices:
\begin{equation} \label{eq:conds}
  \M_{1,2,3} \left( \begin{array}{c} 
    1 \\ 1 \\ 1
  \end{array} \right) = 
  \M_4^T \left( \begin{array}{c} 
    1 \\ 1 \\ 1
  \end{array} \right) = 
  \M_{0,1,4} \left( \begin{array}{c} 
    1 \\ 1 \\  c^2 
  \end{array} \right) =  0\,.
\end{equation}   
{}From these conditions and from the $1 \leftrightarrow 2$ symmetry
it follows that the matrices can be written in terms of  the following 
combinations of projectors
\begin{eqnarray} 
  \M_0 &=&  a_{0} \mathcal{P}_{12} + b_{0} \C^{-2} (\mathcal{P}_{13}+
   \mathcal{P}_{23})\C^{-2}\nonumber \\
  \M_1 &=&  a_{1} \mathcal{P}_{12} \nonumber \\
  \M_2 &=&  a_{2} P_{12} +  b_{2} (\mathcal{P}_{13}+
  \mathcal{P}_{23})
  \nonumber \\
  \M_3 &=& a_{3}  \mathcal{P}_{12} + 
   b_{3} (\mathcal{P}_{13} + 
  \mathcal{P}_{23})\nonumber \\ 
  \M_4 &=&  a_{4} \mathcal{P}_{12} + 
   b_{4} \left(  \mathcal{P}_{13} + \mathcal{P}_{23}\right)\C^{-2} ,
\label{eq:projs}
\end{eqnarray}
where 
{\small
\be
  \mathcal{P}_{12} = \left( \begin{array}{ccc} 
     1 & -1 & 0 \\
     -1 & 1 & 0 \\
     0 & 0 & 0 
  \end{array} \right) \qquad
  \mathcal{P}_{13} = \left( \begin{array}{ccc} 
     1 & 0 & -1 \\ 
     0 & 0 & 0 \\
     -1 & 0 & 1 
  \end{array} \right) \qquad
  \mathcal{P}_{23} = \left( \begin{array}{ccc} 
     0 & 0 & 0 \\
     0 & 1 & -1 \\
     0 & -1 & 1 
  \end{array} \right) \; ,
\ee }
and $a_{i}$, $b_i$ are constants that depend on $V$.

\pagebreak[3]

Since we are interested in the gravitational potential we
focus on the scalar sector. The quadratic  Lagrangian for the scalars  is
\begin{eqnarray} 
  \mathcal{L}_{\text{scalars}}&=& 
 \phi^t \K^2  \Delta \,  \tau 
- \tau^t \frac{\K^2}{2} \left(\C^2 \Delta -3 \de_t^2  
  \right) \tau 
+\nb\\
&&{}+\frac14 \Big[\psi \M_0 \psi -2 \Delta v \M_1 v - 
  (\tau+\Delta \sigma)\M_2 (\tau + \Delta \sigma)-2 \tau \M_2 \tau
\nb\\
  & & \qquad + (3\tau+\Delta\sigma) \M_3
  (3\tau +\Delta\sigma)-2(3\tau+\Delta\sigma) 
  \M_4 \psi \Big]\nb\\
&&{} - \phi\frac{\C^{-3}}{2}  T_{00}  -  \tau^t  \frac{\C}{2} T_{ii}   \,.
\label{eq:L0}
\end{eqnarray}
In order to disentangle the different fluctuations it is useful to 
a `tilded' basis where the fluctuations are rotated:
\newcommand\sq{\frac1{\sqrt{2}}}%
\be
\left [ \psi, v, \sigma , \tau \right] 
= S
\left [ \tilde \psi, \tilde v, \tilde \sigma , \tilde \tau \right]
,    \qquad
\tilde{\M}_i=S^t \M_i S \qquad  
 S=\left( \begin{array}{cccc}
  -\sq & \sq & 0  \\
   \sq & \sq & 0  \\
  0  & 0 & \sq 
  \end{array} \right).
\ee
In this basis the mass matrices take the block-diagonal form
\bea
  \tilde{\M}_0=\left(\! \begin{array}{cccc}
    2a_0+b_0 & 0 & 0 \\
    0 &  b_0   & -b_0 /c^{2} \\
    0 & -b_0 /c^{2} & b_0 /c^{4}    
  \end{array}\! \right), &&\qquad 
  \tilde{\M}_1=  \left(\! \begin{array}{cccc}
     4 a_{1} & 0\ &\ 0\ \\
     0 &0 & 0 \\
    0 & 0 & 0 
  \end{array} \!\right),
\nb\\[1ex]
  \tilde {\M}_{2,3}=\left(\!\!
    \begin{array}{ccc}
      2a_{2,3}+b_{2,3} & 0 & 0 \\
      0 & b_{2,3} & -b_{2,3} \\
      0 & -b_{2,3} & b_{2,3} 
    \end{array} \!\!\right), &&
  \tilde {\M}_{4}=\left(\!\!
    \begin{array}{ccc}
      2a_4+b_4 & 0 & 0 \\
      0 & b_{4} & -b_{4}/c^2 \\
      0 & -b_{4} & b_{4}/c^2 
    \end{array} \!\!\right).
\label{newbasis}
\ena
Because the ${\cal K}$ commutes with $S$, we see that in the new
basis the system naturally splits into two: a single massive gravity
sector and a bigravity sector encoded in  the 2$\times$2 sub-matrices
in~(\ref{newbasis}).  The presence $\bar g_3$ induces in both
sectors a Lorenz breaking  mass pattern. The first sector can be
analysed as in~\cite{rubakov}, while for the second the analysis
of~\cite{pilo} applies.  As a result,  a consistent theory, free of
ghosts and of instabilities at linearized level is possible.
Indeed, in the single massive graviton sector, ghosts can be avoided if
the relevant entry 1-1 in $\tilde{\M}_0$ vanishes. We have thus the
condition:
\be
\label{eq:noghost}
a_{0}=-b_{0}/2\,.
\ee
The bigravity sector on the other hand is automatically free of ghosts
as shown in~\cite{pilo} thanks to the vanishing of $\tilde{\M}_1$ in
the relevant block.

\medskip

At this point we can study in the new basis the static gravitational
potential in each sector captured by  the gauge invariant field
$\tilde{\phi}_a=\tilde{\psi}_a-2\partial_t \tilde{v}_a +\partial_t^2
\tilde{\sigma}_a$ ($a=1,2,3$). It is convenient to define also the
rotated and $M^2$-normalized sources ${\tilde t}_{\m\n}=S
\,t_{\m\n}=S\, (T_{\m\n}/M^2)$.

The field $\tilde{\phi}_1$ is separated from the bigravity sector and
gives the Yukawa-like static potential. It turns out that in general
$\tilde{\phi}_1$ is a combination of two Yukawa potentials, with two
parametrically different mass scales (see appendix~\ref{yuka} for the
details). Here, for notation simplicity, we consider the case where the scales coincide
\begin{equation} 
  \qquad \qquad\tilde{\phi}_1 = \frac{\tilde{t}_1}{2\Delta - m^2}\,, \qquad  
\qquad m^2 = 3(2a_{4}+b_{4})\frac{\epsilon^4}{M^2} \, .
\end{equation}
In this sector, in addition to the propagating massive graviton (two
polarizations) also a vector and a scalar propagate
(respectively two and one degrees of freedom). All these fields are
massive with mass given by the relative 1-1 entry of $\tilde{\M}_2$.
The vector and the scalar can have well behaved properties, i.e.\ no
ghosts when condition (\ref{eq:noghost}) is enforced.
In~\cite{rubakov} it was also argued that the scale of strong coupling
is high enough, coinciding with $\Lambda_2\simeq\sqrt{M m}$, with $m$
the characteristic mass scale in this sector.

For the remaining bigravity sector the gravitational potential can be
computed by solving the equations of motion as in~\cite{pilo}. The result is
\begin{eqnarray} \label{eq:pot_mu}  
    \tilde{\phi}_2 &=& \frac{\tilde{t}_{200}+\tilde{t}_{2iii}}{2\Delta}+
    \mu^2\frac{\tilde{t}_{200}}{\Delta^2} \\
    \tilde{\phi}_3 &=& -\mu^2 \left( \frac{M}{M_3} \right)^2\frac{2 c 
    \omega^2\tilde{t}_{200}}{\Delta^2}
\label{pot23}
\end{eqnarray}
where
\begin{equation}
  \mu^2 = \frac{\epsilon^4}{M^2}\left[b_2\frac{3b_{4}^2+b_{0}(b_{2}-3b_{3})}{2(b_{4}^2
  +b_{0}(b_{2}-b_{3}))} \right] \, .
\end{equation}
When $M_{3}\gg M$, the third sector
has a sub-leading impact on the other gravitational potentials. In the
limit $M_{3} \to \infty $, the third sector decouples and $g_3$ just
produces a LB fixed background $\hat \eta$. Going back
to the original basis, the potentials are:
\be
\begin{split}
  \phi_1 &=  \frac{t_{100}+t_{1ii}+t_{200}+t_{2ii}}{4\Delta} +
     \frac{t_{100}+t_{1ii}-t_{200}-t_{2ii}}{4\Delta-2m^2} +
     \mu^2 \frac{t_{100}+t_{200}}{2 \Delta^2} \\[1ex]
  \phi_2 &=   \frac{t_{100}+t_{1ii}+t_{200}+t_{2ii}}{4\Delta} -
     \frac{t_{100}+t_{1ii}-t_{200}-t_{2ii}}{4\Delta-2m^2} +
     \mu^2 \frac{t_{100}+t_{200}}{2 \Delta^2}\\[1ex]
  \phi_3 &=  -\mu^2 \left( \frac{M}{M_3} \right)^2\frac{c 
    \omega^2 (t_{100}+t_{200})}{\Delta^2}\,.  
\end{split}
\label{bpot}
\ee 
The potentials $\phi_{1,2}$ contain a Newtonian term, a
Yukawa-like term, and a linearly growing term, originating from
$\m^2/\Delta^2$.

This latter linear term is the same appearing in the bigravity case,
as found in~\cite{pilo,dubovsky01}. It would invalidate perturbation
theory at distances larger than $r_{IR}^{-1}\sim G \m^2 M_\ast$ from a
source $M_\ast$~\cite{pilo}, but remarkably the full nonlinear
solutions found in~\cite{pilo1} shows that its linear growth is
replaced by a non-analytic power $\sim r^\gamma$, where $\gamma$
depends on the coupling constants in the potential.  Moreover, in the
full solution for a realistic star, also the magnitude of this new
term is proportional to $\m^2$, therefore the effect can be eliminated
by setting $\m^2=0$.  This can be achieved by simple fine-tuning, or
by adopting a particular scaling symmetry of the potential, as
discussed in~\cite{pilo}. We can thus obtain a pure Yukawa
modification of the gravitational potential, by setting $\m^2=0$, that
here amounts to the condition $b_{0}=-3b_{4}^2/(b_{2}-3b_{3})$.

\medskip

The analysis of vector modes is identical to that carried out
in~\cite{rubakov} for the single gravity sector and to the one
of~\cite{pilo} for the bigravity one. In the single-gravity sector
there is a vector state propagating with a nonlinear dispersion
relation: at high energy its speed is $(2a_2+b_2)/(2a_1+b_1)$ and at
low momentum it has a mass gap given by $b_2/M^2$. In the bigravity
sector vector states do not propagate.

\medskip

\newcommand\BOX{\rule{0pt}{1.4ex}\Box}%

The analysis of tensor modes is similar and is best carried out in the
original basis. In the limit of $M_3\to\infty$, the equation of motion
for the canonically normalized fields becomes:
\be
\left[\PM{\Box \\&\!\!\Box\\&&\!\!\widehat\BOX}+\frac1{M^2}\PM{b_2+a_2&b_2-a_2&0\\b_2-a_2&b_2+ a_2&0\\0&0&0}\right]\PM{\chi^c_{1\,ij}\\\chi^c_{2\,ij}\\\chi^c_{3\,ij}\\}
=\PM{t_{1\,ij}\\t_{2\,ij}\\0}
\ee
where $\Box=\eta^{\m\n}\partial_\m\partial_\n$,
$\widehat{\BOX}=\hat\eta^{\m\n}\partial_\m\partial_\n$ and we used the
form of the projectors~(\ref{eq:projs}). We see that the massless spin
two state decouples (it is a superimposition of mostly $\chi_3$) and
we are left with two massive gravitons, with two polarizations each,
travelling at the normal speed of light.  Their mass matrix can be
diagonalized, and the resulting graviton masses are
$m_{g1}^2=(2a_2+b_2)\epsilon^4/M^2$, $m_{g2}^2=b_2\epsilon^4/M^2$.

\medskip

As an explicit example, consider the simplest case of a potential
quadratic in $X_{12}$, $X_{13}$, $X_{23}$ plus two cosmological terms, 
satisfying the $1
\leftrightarrow 2 $ exchange symmetry (taking, for simplicity 
$\omega=1$):
\begin{eqnarray} 
  V(g_1,g_2,g_3)\!\! &=\!\!& \xi_0 + \xi_1 \left( {\rm Tr}[X_{13}^2] + 
  {\rm Tr}[X_{23}^2] \right) +
  \xi_2 \, {\rm Tr}[X_{13} X_{23}]  + 
  \nonumber \\   
  & & \xi_3 \left( {\rm Tr}[X_{13}]^2 + {\rm Tr}[X_{23}]^2 \right) +
  \xi_4 \, {\rm Tr}[X_{13}] {\rm Tr}[X_{23}] + 
  \nonumber  \\
   & & \xi_5 \left( {\rm Tr}[X_{12}]^2 + {\rm Tr}[X_{12}^{-1}]^2 \right) +
  \xi_6 \left( {\rm Tr}[X_{12}^2] +{\rm Tr}[(X_{12}^{-1})^2] \right) + 
  \nonumber \, \\
  & & \xi_7 \left( ({\rm det}X_{12})^{-1/6}  ({\rm det}X_{13})^{-1/6} + 
  ({\rm det}X_{12})^{1/6}  ({\rm det}X_{23})^{-1/6} \right) + \nonumber \\
  & & \xi_8 ({\rm det}X_{13})^{1/6}  ({\rm det}X_{23})^{1/6}
  \label{eq:qpot}
\end{eqnarray}
The EOMs for a flat background require to solve for three constants
(e.g.  $\xi_3,\xi_7,\xi_8$) and then the coefficients of the
projectors in the mass matrices $a_i$'s and $b_i$ are a function of the
remaining coupling constants (see appendix~\ref{p_fun}).

The no-ghost condition $b_{0}=-2a_{0}$, the condition for the absence
of the linear term $\mu^2=0$ and the condition for having a single
Yukawa scale (see Appendix \ref{yuka}), can be solved for
$\xi_{1,2,4}$ and we end up only four
couplings eventually.  The Yukawa scale $m$ and the graviton masses
$m_{g1}^2$ and $m_{g2}^2$, only depend on $\xi_5$ and $\xi_6$:
\begin{eqnarray}
	m^2 &=& [p_0(c) \xi_5 + q_0(5) \xi_6]\frac{\epsilon^4}{M^2} \nonumber \\
\label{eq:gmasses}
   m_{g1}^2 &=& [p_1(c) \xi_5 +q_1(c) \xi_6]\frac{\epsilon^4}{M^2}  \\
   m_{g2}^2 &=& [p_2(c) \xi_5 +q_2(c) \xi_6]\frac{\epsilon^4}{M^2} \nonumber
\end{eqnarray}  
where $p_i(c)$'s and $q_i(c)$'s are given in
appendix~\ref{p_fun}.\footnote{When $\xi_6=0$ and all the masses above
  depend only on $\xi_5$, one can check that they are positive, for $
  1.41 \lesssim c \lesssim 2.05$.}

%

To summarize, in the limit where the third metric is decoupled the
theory has two massive gravitons and the potential felt by a
test particle of type 1 is: 
\begin{equation} 
  \phi_1(r) = \frac{G\, m_1}{r} \left(\frac{1+ 
  e^{-mr}}{2} \right) +
              \frac{G\, m_2}{r} \left(\frac{1-  
  e^{-mr}}{2} \right),   
\label{rotpot}
\end{equation}
where $G=1/16\pi M^2$. This shows that the vDVZ discontinuity is absent,
and we have obtained the potential~(\ref{bigr}) while avoiding the
troubles of the Lorentz-invariant FP theory.

\section{Conclusions}

Motivated by the interesting possibility to relax the assumption that
dark and visible matter feel the same gravitational interaction, in
this work we have addressed the possibility to obtain a Yukawa-like
large-distance modification of the gravitational potential, while
avoiding ghosts or classical instabilities. 

The request to  generate an Yukawa potential from a consistent theory
led us to consider Lorenz-Breaking backgrounds in a suitable extended class of
bigravity theories.  For instance, bigravity  while giving rise to a
healthy Lorentz-Breaking massive phase, does not produce a Yukawa
potential. Here we have generalized this picture and have shown that
if an additional field $g_3$ is introduced, a Yukawa modification is
allowed. The extra field $g_3$ can be harmlessly decoupled by freezing
it to a Lorentz violating background configuration.  The two remaining
sectors represent two interacting massive gravities, of which one
features a Yukawa potential.  This pattern then leads to the desired
modified gravity where standard matter (type 1) couples to all the
graviton mass eigenstates.

On the technical side, the price to be paid to solve the problem is
that two fine-tunings are needed, one to have a ghost-free spectrum at
linear level, the other to avoid the linearly growing potential.  The
first one has be shown to follow (in single massive gravity theories)
from extra unbroken partial diffeomorphisms invariance~\cite{rub_rev}
and it would be interesting to extend that symmetry arguments also to
the present model. The other can also be understood as the consequence
of a scaling symmetry of the potential~\cite{pilo}. 

Let us note also that the theory presented here has three rank-two
fields but only nine polarizations propagate (and are well behaved at
quadratic level: three spin-2 with two polarizations each, one spin-1
with two polarizations and one scalar). On the other hand one may
expect that the total number of propagating modes would be 22 = 3\x10
- 4\x2 (twice the number of diagonal diffeomorphisms), and the nine
missing modes will probably propagate at non-linear level. The real
non-perturbative question, to be addressed in a future work, is then
at which scale these non-linear effects would show up.

The resulting setup, featuring two separate metric fields for the
visible and dark matter, allows to consider also collisional and
dissipative dark matter, as mirror matter and the potential generated
by the ordinary and dark matter sources of mass $m_1$, $m_2$ felt by
ordinary matter is distance dependent as
in~(\ref{rotpot}).\footnote{Let us remark also that the weak
  equivalence principle does not exclude the possibility of direct
  (non-gravitational) interactions
  between the normal (type 1) and dark (type 2) matter components. To
  the action (\ref{bg}), besides the interaction gravitational term
  $V$, the mixed matter term $ \int d^4x (g_1 g_2)^{1/4} {\cal L}_{\rm
    mix}$ can be added with the Lagrangian ${\cal L}_{\rm mix}$
  including for example, the photon kinetic mixing term $\varepsilon
  F^{\mu\nu}_1 F_{2\mu\nu}$ \cite{Holdom} or the neutrino interaction
  terms \cite{FV}. This also makes possible the direct detection of
  dark matter via such interactions \cite{Foot}, with interesting
  implications. } The result is very different from the standard
picture when normal and dark matters both feel an universal Newtonian
gravity; in fact the potential~(\ref{rotpot}) can be used to fit
galactic rotational curves using similar density profiles for the
visible and dark sectors, alleviating the problems of profile
formation ~\cite{curves}. 

In order to grasp the basic idea take, for instance,
a spiral galaxy made of two overlapped disks of type~1 and type~2 matter. 
Both types of matter are distributed
along the  disks according to a similar density
profile $\sigma_{1,2}(r)=\sigma_{1,2} e^{-r/r_{1,2}}$, where
$\sigma_{1,2}$ are the central densities and $r_{1,2}$ are
the core radii of the two types of matter respectively. The
velocity profile is given by $v(r)=\sqrt{a(r)r}$, where 
$a(r)$ is the centrifugal acceleration of a star rotating
at distance $r$ from the center, obtained by integrating
over the disks the  acceleration $g(|({\bf r}- {\bf r'}|)$ due to a  point-like source
\begin{equation}
\begin{split}
 a(r) &= \int_{\text{disk}}  d^2r' g({\bf r}- {\bf r'})
 \frac{{\bf r}-{\bf r'}}{|{\bf r}-{\bf r'}|} \, , \\
 g({\bf r}- {\bf r'})
 &=\frac{G}{2}\frac{\left(
 \sigma_1(r')+\sigma_2(r')\right)}{|{\bf r}- {\bf r'}|^2}+\frac{G}{2}
 \frac{\left(\sigma_1(r')-\sigma_2(r')\right)}{|({\bf r}- 
  {\bf  r'}|^2}\left(1+\frac{|({\bf r}- {\bf r'}|}{r_m}\right)
  e^{|({\bf r}- {\bf r'}|/r_m}. 
\end{split}
\end{equation}
From this expression, one can see that in a wide region
around $r_m$ the gravitation interaction between the two types of matter turns on enhancing the rotational velocity over the Keplerian fall which would have given $v(r)\propto r^{-1/2}$.  
At large distance $r\gg r_m$, the behaviour is once again
Keplerian-like, but with a crucial difference that the force is due not only to the visible matter, but the
total mass average $(M_1+M_2)/2 > M_1$.
It is interesting to note that the effective
Newton constant relative to the type~1~-~type~1 and type~2~-~type~2
interactions is distance dependent: $G_N(r \ll m^{-1})=G$ and $G_N(r
\gg m^{-1})=G/2$.

Finally, let us  comment on the cosmological implications of our model. Though a detailed study is left for a future work, we do not expect a strong impact at cosmological distances;
for instance, the modified potential and the presence of a new type of matter will not change the expansion of the universe. The scale of gravity modification is set by the inverse of the graviton mass, which is about 10 kpc in our model. As a result, at cosmological scales  only the Newtonian-like mode is active with an halved Newton constant $G/2$ and type 2 matter should behave as cold dark matter. The observed Hubble parameter would imply for the total density of the universe a value twice bigger being the effective critical density halved, e.g. $\rho_{cr} =3 H_0^2/ 4 \pi G$ instead of $\rho_{cr} =3 H_0^2/ 8 \pi G$. Then to reproduce the ratio $\Omega_B/\Omega_{DM}$, the mirror matter density should be about 10 times bigger than the baryon density.

In conclusion, at linearized  level a Yukawa modification of the
Newtonian gravitational potential is possible and it  opens up
the possibility to have collisional dark matter, coupled to ordinary
matter via a modified gravitational interaction in a physically
nontrivial way.



\section*{Acknowledgments} 

We thank D. Comelli for useful discussions.  The work is supported in
part by the MIUR grant for the Projects of National Interest PRIN 2006
"Astroparticle Physics", and in part by the European FP6 Network
"UniverseNet" MRTN-CT-2006-035863.

\begin{appendix}

\section{General Yukawa-like potential}
\label{yuka}

The degrees of freedom in the massive gravity sector consists of a
metric fluctuation with mass term that we can write as
\begin{equation} \label{eq:mass_r}
    \mathcal{L}_{\text{mass}} = \frac{M^2}{2} \left(
   m_0^2 h_{00} h_{00} + 2 m_1^2 h_{0i} h_{0i} - m_2^2 h_{ij} h_{ij} + 
   m_3 h_{ii} h_{jj} - 2 m_4^2 h_{00} h_{ii} \right).
\end{equation} 
In our case effectively $m_0=0$ and when $m_1 \neq 0 $,
there is a healthy propagating scalar degree of freedom ($\tau$) as
well as a healthy propagating vector~\cite{rubakov}. The scalar
perturbations obey the equations:
\begin{eqnarray}
 &&   2 \Delta \tau - m_4^2(\Delta \sigma + 3 \tau) = t_{00} \\
 &&   2 \partial_0 \tau - m_1^2 v  = \frac{1}{\Delta} \partial_0 t_{00} \\
 &&  2 \partial_0^2 \tau - m_2^2 \Delta\sigma - m_2^2 \tau + m_3^2\Delta \sigma
  + 3 m_3^2 \tau - m_4^2 (\phi + 2 \partial_0 v - \partial_0^2 \sigma) = 
  \frac{1}{\Delta} \partial_0^2 t_{00} \\
 &&  2 \Delta \phi - 2 \Delta \tau + 2 m_2^2 \Delta \sigma = 
  t_{ii} - \frac{3}{\Delta} \partial_0^2 t_{00}\,,    
\end{eqnarray}
where $ t_{\mu\nu} = T_{\mu\nu}/M^2 $.  The equations can  be solved for
$\phi$ to get the static gravitational potential. One finds
\begin{equation}
  \phi = \frac{(t_{00} + t_{ii})(\zeta_1 - 1)\zeta_2\Delta + 
  [t_{ii} + t_{00}(3 \zeta_1 - 1)\zeta_2]\zeta_2m_4^2}{2 (\zeta_1-1)\zeta_2 \Delta^2 +
  (4 \zeta_2 - 1) m_4^2 \Delta -3 \zeta_2 m_4^4}\,,
\end{equation}
with $ \zeta_1 = m_3^2 / m_2^2 $ and $ \zeta_2 = m_2^2 / m_4^2 $. 
This potential can be written as the sum of two Yukawa-like  terms:
\begin{equation}
  \phi = \frac{t_+}{2(\Delta - m_+^2)} +
  \frac{t_-}{2(\Delta - m_-^2)}\,, 
\end{equation}
where
\bea
   t_\pm &=& \frac{1}{2} \left[ 1\pm \frac{1}{2} \left( \nu +
   \frac{1}{\nu} \right) \right] t_{00} + \frac{1}{2}  
   \left(1\pm \frac{1}{\nu} \right)t_{ii} \, , \\  
m_\pm^2 &=& m^2_4 \frac{6 \zeta_2}{4\zeta_2-1\pm\nu} ,
\qquad\qquad    
\nu = \sqrt{1+8\zeta_2(3 \zeta_1\zeta_2-\zeta_2-1)} \, .
\nonumber
\ena

Recall~\cite{rubakov} that the conditions $ \zeta_2 > 1/4 $ and $
\zeta_1 < 1 $ ensure that the theory has no derivative instabilities
neither in the UV nor in the IR. Moreover, if
$\zeta_1>(8\zeta_2^2+8\zeta_2-1)/24\zeta_2^2$, 
then $m_{\pm}^2$ are real and positive, and the theory has no instabilities
also at intermediate scales. Accordingly, the potential is  the
sum of two ``genuine'' Yukawa-like terms.
Finally, if $\zeta_1 = (1+\zeta_2) / 3 \zeta_2 $, then $\nu=1$ and
$t_-$ vanishes, so that one is left with a single Yukawa potential:
\begin{equation}
  \phi = \frac{t_{00} + t_{ii}}{2 (\Delta - \frac{3}{2} m_4^2)}\,.
\end{equation}

\section{Explicit solution for potential~(\ref{eq:qpot})}
\label{p_fun}

After solving the EOMs (\ref{eq:const}), the coefficients that enter the masses for potential~(\ref{eq:qpot})  can be written in terms of the coupling constants and read
\begin{eqnarray} \label{eq:ab_par}
   a_0 &=& -\frac{\xi_4 c^4}{2}-\frac{2 \xi_1
   c^2}{9}-\frac{\xi_0}{72}-\frac{c^2}{18} \left(9 c^2+2
   \right) \xi_2+\frac{35 \xi_5}{9}+\frac{50
   \xi_6}{9}\nonumber \\
	b_0 &=& -\frac{\xi_0}{72}-\frac{c^2\left(39 -23 c^2\right)
   \xi_1}{18 \left(c^2+3\right)}-\frac{c^2\left(39 -23
   c^2\right) \xi_2}{36 \left(c^2+3\right)}-\frac{\xi_5}{9}-\frac{\xi_6}{9}
\nonumber \\
	a_1 &=& \frac{c^2 \xi_2}{2}-4 \xi_5 \nonumber \\
	a_2 &=& \frac{\xi_2}{2}-\xi_5 \nonumber \\
	b_2 &=& \left(c^2-1\right) \xi_1+\frac{1}{2} \left(c^2-1\right)
   \xi_2\nonumber \\
   a_3 &=& -\frac{2 \xi_1 c^2}{9}-\frac{\xi_2
   c^2}{9}-\frac{\xi_0}{72}-\frac{
   \xi_4}{2}+\frac{26 \xi_5}{9}+\frac{50 \xi_6}{9}\\
   b_3 &=& {}-\frac{\xi_0}{72}+\frac{\left(5 c^2-6 c\right)
   \xi_1}{18}+\frac{\left(5 c^2-6
   \right) \xi_2}{36}-\frac{\xi_5}{9}-\frac{4 \xi_6}{9} \nonumber\\
   a_4 &=& -\frac{2 \xi_1 c^2}{9}-\frac{\xi_2
   c^2}{9}-\frac{\xi_4 c^2}{2}-\frac{\xi
   _0}{72}-\frac{\xi_5}{9}+\frac{50 \xi_6}{9}\nonumber \\
   b_4 &=& -\frac{\xi_0}{72}-\frac{\left(13 c^4+3 c^2\right)
   \xi_1}{18 \left(c^2+3\right)}-\frac{\left(13 c^4+3
   c^2\right) \xi_2}{36 \left(c^2+3\right)}-\frac{\xi_5}{9}-\frac{4\xi_6}{9} \nonumber 
  \nonumber 
\end{eqnarray}

Finally, the functions $p$ appearing in the graviton masses~(\ref{eq:gmasses}) can be written as
\begin{eqnarray}
p_0 &=& \frac{1}{C_2} \left[ 6 \left(c^2 \left(\left(3 \left(1850 c^8-7725 c^6-31099
   c^4+154507 c^2-92547\right) c^2+7 C_1-   
   \right. \right. \right.\right. \nonumber \\ 
   & & \left. \left. \left. \left. 168318\right) c^2 -5 C_1+117936\right)-
   18 C_1\right) \right] \\
q_0 &=& \frac{1}{C_2} \left[ 12 \left(c^2 \left(\left(3610 c^{10}-15312 c^8-58045
   c^6+296415 c^4-187461 c^2+ \right. \right. \right. \right. \nonumber \\
	& & \left. \left. \left. \left. 
   14 C_1-115371\right) c^2-10
   C_1+88452\right)-36 C_1\right) \right] \\
p_1 &=& -\frac{1}{C_3} \left[ 2 \left(95 c^{18}-5674 c^{16}+21090 c^{14}+95053
   c^{12}-447746 c^{10}+ \right. \right.  \nonumber \\
   & & \left. \left.
 243567 c^8+194157 c^6-37
   C_1\left( 6 c^8-125117 c^4+5  c^2+42\right)
   \right) \right] \\
q_1 &=& -\frac{1}{C_3} \left[ 2 \left(380 c^{18}-7716 c^{16}+22284 c^{14}+123114
   c^{12}-529730 c^{10} + 334098 c^6-148  + \right. \right.  \nonumber \\
   & & \left. \left.    C_1 \left( 24 c^8  -205631c^4+20 c^2+168
   238626 \right) 
   \right) \right] \\
p_2 &=&-\frac{\left(95 c^{10}-324 c^8-405 c^6+378 c^4+6 C_1\right)
   C_4}{C_5}\\
q_2 &=&-\frac{4 \left(95 c^{10}-324 c^8-405 c^6+378 c^4+6 C_1\right)
   C_4}{C_5}\,,
\end{eqnarray}
where
\begin{eqnarray}
C_1  &=& c^4 \left(5 c^4-26 c^2+21\right) \sqrt{13 c^4+42 c^2+9} \\
C_2 &=& (1404 - 1773 c^2 + 6 c^4 + 107 c^6) (c^4 (-2 + c^2) (-21 + 5 c^2))\\
C_3 &=& c^8 (58968 - 117990 c^2 + 62235 c^4 - 4557 c^6 - 3287 c^8 + 535 c^{10})\\
C_4 &=& 2 \left(c^4+2 c^2-3\right)\\
C_5 &=& c^6 (-29484 + 44253 c^2 - 8991 c^4 - 2217 c^6 + 535 c^8)\,.
\end{eqnarray}

\end{appendix}



\begin{thebibliography}{99}
  

\bibitem{FP} 
  M.~Fierz and W.~Pauli,
  Proc.\ Roy.\ Soc.\ Lond.\  A {\bf 173}, 211 (1939).

\bibitem{DIS} 
  H.~van Dam and M.J.G.~Veltman,
  Nucl.\ Phys.\  B {\bf 22}, 397 (1970); \\
  Y.~Iwasaki,
  Phys.\ Rev.\  D {\bf 2}, 2255 (1970); \\
  V.I.~Zakharov,
  JETP Lett.\  {\bf 12}, 312 (1970) 
  [Pisma Zh.\ Eksp.\ Teor.\ Fiz.\  {\bf 12}, 447 (1970)].
 

\bibitem{boul_des} 
  D.G.~Boulware and S.~Deser,
  Phys.\ Rev.\  D {\bf 6}, 3368 (1972).

\bibitem{NGS} 
  N.~Arkani-Hamed, H.~Georgi and M.D.~Schwartz,
  Annals Phys.\  {\bf 305}, 96 (2003) 
  [hep-th/0210184].

\bibitem{ghostcond} 
  N.~Arkani-Hamed, H.C.~Cheng, M.~Luty and J.~Thaler,
  JHEP {\bf 0507}, 029 (2005)
  [hep-ph/0407034].

\bibitem{rubakov}
  V.A.~Rubakov,
  [arXiv:hep-th/0407104].

\bibitem{dubos}
  S.~L.~Dubovsky,
  JHEP {\bf 0410}, 076 (2004)
  [arXiv:hep-th/0409124].

\bibitem{kb} 
  R.~Bluhm and V.~A.~Kostelecky,
  Phys.\ Rev.\  D {\bf 71} (2005) 065008
  [arXiv:hep-th/0412320].

\bibitem{Isham}
  C.J.~Isham, A.~Salam and J.A.~Strathdee,
  Phys.\ Rev.\  D {\bf 3}, 867 (1971);
  T.~Damour, I.I.~Kogan,
  Phys.\ Rev.\  D {\bf 66}, 104024 (2002).

\bibitem{pilo} 
  Z.~Berezhiani, D.~Comelli, F.~Nesti and L.~Pilo,
  Phys.\ Rev.\ Lett.\  {\bf 99}, 131101 (2007)
  [hep-th/0703264]; 

\bibitem{Blas:2007zz}
  D.~Blas, C.~Deffayet and J.~Garriga,
  Phys.\ Rev.\  D {\bf 76} (2007) 104036
  [arXiv:0705.1982 [hep-th]].



\bibitem{pilo1}
 Z.~Berezhiani, D.~Comelli, F.~Nesti and L.~Pilo,
  JHEP {\bf 0708}, 130 (2008)  [arXiv:0803.1687  hep-th].

\bibitem{sal1}
P.~Salucci, A.~Lapi, C.~Tonini, G.~Gentile, I.~Yegorova and U.~Klein,
Mon.\ Not.\ Roy.\ Astron.\ Soc.\  {\bf 378} (2007) 41
[arXiv:astro-ph/0703115].

\bibitem{sal2}
P. Salucci, F. Walter and A. Borriello, A\&A (2003),  409, 53; \\
G. Gentile, A. Burkert, P. Salucci, U. Klein and F. Walter, ApJ (2005),
634, 145

\bibitem{mirror}
I.Yu.\ Kobzarev, L.B.\ Okun and I.Ya.\ Pomeranchuk,  
Sov.\ J.\ Nucl.\ Phys. 3, 837 (1966); \\
S.G.\ Blinnikov and M.Yu.\ Khlopov, Sov.\ Astron. 27, 371 (1983); \\ 
R.\ Foot, H.\ Lew and R.R.\ Volkas, Phys.\ Lett.\ B 272, 67 (1991). \\
For asymmetric case, see  
Z. Berezhiani, A. Dolgov and R.N. Mohapatra, 
Phys. Lett. B {\bf 375}, 26 (1996); 
Acta Phys. Pol. B {\bf 27}, 1503 (1996).
 
\bibitem{alice} 
Z. Berezhiani,  Int. J. Mod. Phys. A19, 3775 (2004); 
Eur.  Phys. J. ST {\bf 163}, 271 (2008);  
  AIP Conf.\ Proc.\  {\bf 878} (2006) 195; 
hep-ph/0508233


\bibitem{BB-PRL}
L. Bento and Z. Berezhiani, Phys. Rev. Lett. {\bf 87}, 231304 (2001); 
Fortsch. Phys. {\bf 50}, 489 (2002); 
hep-ph/0111116 


\bibitem{Holdom}
B. Holdom, Phys. Lett. B {\bf 166}, 196 (1986); \\
S.L. Glashow, Phys. Lett. B {\bf 167}, 35 (1986). 


\bibitem{FV}
R. Foot and R.R.\ Volkas, Phys. Rev. D {\bf 52}, 6595 (1995); \\
Z. Berezhiani and R.N. Mohapatra, Phys. Rev. D {\bf 52}, 6607 (1995); \\
Z. Berezhiani and L. Bento, Phys. Rev. Lett. {\bf 96}, 081801 (2006)    
[hep-ph/0507031] 

\bibitem{BCV}
Z. Berezhiani, D. Comelli and F. Villante,
Phys. Lett. B {\bf 503}, 362 (2001);  \\
%
A.Yu.\ Ignatiev and R.R. Volkas, Phys. Rev. D {\bf 68}, 023518 (2003); \\ 
Z.~Berezhiani, P. Ciarcelluti, D. Comelli and F. Villante, 
Int.\ J.\ Mod.\ Phys.\ D {\bf 14}, 107 (2005). 


\bibitem{curves} 
Z. Berezhiani, L. Pilo, N. Rossi, arXiv:0902.0146[hep-ph]

\bibitem{Banados:2008fj}
  M.~Banados, P.~G.~Ferreira and C.~Skordis,
 Phys.\ Rev.\  D {\bf 79}, 063511 (2009)
  [arXiv:0811.1272 [astro-ph]].


\bibitem{will} 
C.M. Will, arXiv:gr-qc/0510072. 

\bibitem{Kancheli} 
Z. Berezhiani and O.V. Kancheli, arXiv:0808.3181 [hep-th]


\bibitem{dubovsky01} 
  S.L.~Dubovsky, P.G.~Tinyakov and I.I.~Tkachev,
  Phys.\ Rev.\ Lett.\  {\bf 94},  181102 (2005)

\bibitem{rub_rev}
V.A.~Rubakov and P.G.~Tinyakov,
Phys.\ Usp.\  {\bf 51}, 759 (2008) 
[arXiv:0802.4379 [hep-th]].

\bibitem{Foot} 
R. Foot, Phys. Rev. D78, 043529 (2008),  arXiv:0804.4518 [hep-ph] 






		        
\end{thebibliography}
\end{document}